\documentclass[useAMS,usenatbib]{mn2e}
\usepackage{epsfig}
\usepackage{epstopdf}
\usepackage{graphicx,amssymb,supertabular,longtable,lscape} 
\usepackage{amsmath}
\usepackage{textcomp}

\def\apj{\mbox{ApJ}}
\def\apjl{\mbox{ApJL}}
\def\apjs{\mbox{ApJS}}

\def\mnras{\mbox{MNRAS}}

\def\nat{\mbox{Nature}}
\def\aap{\mbox{A\&A}}

\def\nar{\mbox{New Astronomy Reviews}}

\def\mum{\textmu m }
\def\mums{\textmu m}

\title[Smooth and Clumpy dust distribution in AGN]
{Smooth and Clumpy Dust Distribution in AGN: a Direct Comparison of two Commonly Explored Infrared Emission Models}

\author[Feltre et al.]{A. Feltre$^{1,2}$\thanks{e-mail: feltre.anna@gmail.com}, E. Hatziminaoglou$^{2}$, J. Fritz$^{3}$, A. Franceschini$^{1}$\\
$^{1}$Dipartimento di Fisica e Astronomia, Vicolo Osservatorio 2, I-35122 Padova, Italy\\
$^{2}$ESO, Karl-Schwarzschild-Str. 2, 85748 Garching bei M\"unchen, Germany\\
$^{3}$Sterrenkundig Observatorium Vakgroep Fysica en Sterrenkunde Universeit Gent Krijgslaan 281, S9  9000 Gent}
\begin{document}

\pagerange{} \pubyear{0000}

\maketitle

\label{firstpage}

\begin{abstract}

The geometry of the dust distribution within the inner regions of Active Galactic Nuclei (AGN) is still a debated issue and relates directly with the AGN unified scheme. Traditionally, models discussed in the literature assume one of two distinct dust distributions in what is believed to be a toroidal region around the Supermassive Black Holes: a continuous distribution, customarily referred to as {\it smooth}, and a concentration of dust in clumps or clouds, referred to as {\it clumpy}.

In this paper we perform a thorough comparison between two of the most popular models in the literature, namely the smooth models by \cite{fritz06} and the clumpy models by \cite{nenkova08a}, in their common parameters space. Particular attention is paid to the silicate features at $\sim$9.7 and $\sim$18 \mums, the width of the infrared bump, the near-infrared index and the luminosity at 12.3 \mums, all previously reported as possible diagnostic tools to distinguish between the two dust distributions. We find that, due to the different dust chemical compositions used in the two models, the behaviour of the silicate features at 9.7 and 18 \mum is quite distinct between the two models. The width of the infrared bump and the peak of the infrared emission can take comparable values, their distributions do, however, vary. The near-infrared index is also quite different, due partly to the primary sources adopted by the two models. Models with matched parameters do not produce similar SEDs and virtually no random parameter combinations can result in seemingly identical SEDs.

\end{abstract}


\section{INTRODUCTION}\label{intro}

Active galactic nuclei (AGN) can be classified in a variety of ways depending on the characteristics of their spectral energy distribution (SED) at various wavelengths. The commonly used division in type 1 and type 2 AGN is primarily based on their properties in the UV/optical wavelengths, with type 1 objects typically showing broad emission lines, while type 2 only having narrow emission lines in their spectra.
According to the unified scheme for AGN \citep{antonucci93,urri95}, the differences between the type 1 and type 2 AGN are an orientation effect, as first suggested by \cite{rees69}. For certain lines of sight the dust can obscure the central engine giving rise to a number of differences in the observed SED of AGN at almost all wavelengths. This happens as dust grains, present in an optically thick region surrounding the nucleus, absorb ultraviolet photons coming from the central region and re-radiate them in the infrared (IR). Therefore, the presence of dust around the central region of AGN is the key to understanding the  differences between type 1 and type 2 objects.

The variety of radiative transfer models developed to reproduce the observed dust emission in the IR can be divided in two classes: ``smooth'' models characterised by a continuous dust distribution in a toroidal or flared-disk shape in which the density can only vary smoothly within the torus, and ``clumpy'', in which the dust is distributed in clumps or clouds. Clumpy models are a more likely representation of the real dust distribution as a smooth dust distribution would result in collisions that would raise the temperature to levels too high for the dust to survive 
(see e.g. \citealt[][]{krolik88}). 
On the other hand, smooth models were the first to be developed, being computationally simpler and, in many aspects, a good approximation when calculating the IR SED of AGN.

Smooth models, however, confront important issues when used to match the observations: the IR bump they produced was narrower than what was observed  \citep[][see also \S \ref{sec:IRSED}]{dullemond05}; the 10 \mum silicate feature (\S \ref{sec:sifeature}) was often observed in absorption in type 2 sources, but had barely been seen in emission in either type 1 or type 2 sources \citep{dullemond05} until recent observations with {\it Spitzer}/IRS; and the same feature in absorption in type 2 views had always been observed to be shallower than what the model predicted.
With this in mind, \cite{nenkova02} were the first to present a torus model with dust distributed in clumps, with the silicate emission feature attenuated and the IR bump broad enough to fit the observations. The same models, however, fail to reproduce the short wavelength emission emerging from the hot dust \citep[e.g.][]{mor09} in type 1 AGN. This behaviour of the clumpy models is not specific to this particular set of models but appears in other clumpy model realisations \citep[e.g.][]{polletta08}. 

The success of both classes of models in fitting different parts of the observed AGN SEDs keeps the issue of the dust distribution in AGN open, as no conclusions can be drawn from the simple comparison between observed and model SEDs. The differences arising from the two geometries have so far been the focus of two different studies. \cite{dullemond05} compared their own two-dimensional radiative transfer models of smooth and clumpy tori in terms of the resulting width of the SED, strength of the silicate feature at 9.7 \mum and isotropy parameter. They concluded that, despite the distinct nature of the models and the variations this may cause to the shape of the SEDs, distinguishing between the two distributions based on the broad band SEDs was not possible. \cite{schartmann08} implemented a three dimensional clumpy model and compared it with their previous continuous model \citep{schartmann05} as well as other clumpy configurations. Their analysis, with an emphasis on the behaviour of the 9.7 \mum silicate feature, confirmed their previous conclusion that the mid-IR SEDs of AGN are mainly determined by the innermost part of the torus. More recently, \cite{stalevski12} explored the implications of clumpy and continuous dust distributions on the IR SEDs, by exploiting the three-dimensional radiative transfer code SKIRT \citep{baes11}, and found globally no significant dissimilarities in their full set of models.

Other than the works mentioned above, a systematic comparison between the models currently used in the literature is missing. The work we present here intends to partly fill this gap by providing a thorough comparison between the two probably most used smooth and clumpy models in the literature to date, namely an updated grid of smooth models based on Fritz, Franceschini \& Hatziminaoglou (2006) (see \S \ref{sec:smooth} for the details) and the updated clumpy model grid by \cite{nenkova08a}. The two sets of models and the parameter space they cover are described very briefly before constructing restricted grids of the two classes of SEDs by matching the model parameters (\S \ref{sec:models}). We then characterize and compare the two sets of SEDs and
investigate the derived distributions of the respective model parameters (\S \ref{sec:comparison}). We finally summarise our results and expose the implications for the use of the two approaches of radiative transfer models to reproduce the observed AGN SEDs (\S \ref{sec:discussion}).

\section{DESCRIPTION OF MODELS AND GRIDS SELECTION}\label{sec:models}
In this section we briefly summarize the main features of the models compared in this work, describing their main features and parameters, as well as the selection of the final grids of models on which we perform the comparison. For the models details we refer to  \cite{fritz06} and  \cite{nenkova08a,nenkova08b}, hereafter F06 and N08, respectively. What follows is a brief description of the two different model grids. For a complete and detailed description see the respective papers.

\subsection{Smooth dust distribution}\label{sec:smooth}
The grid of smooth models used for this study is based on the F06 models, with some minor updates, as described below. The original models are among the most popular smooth models in the literature to date \citep[see e.g.][]{rodighiero07,berta07,hatzimi08,hatzimi09,hatzimi10,agol09,pozzi10,natale10,vignali09,vignali11}. 
The mixture of graphite (53\%) and silicate grains (47\%), the distribution of grain sizes (Mathis, Rumpl \& Nordsieck 1997) and absorption and scattering coefficients (taken from \citealt{laor93}) remain unchanged with respect to F06. The spectral index for the power laws describing the central source has been updated, following \cite{schartmann05}:  
\begin{equation}
L (\lambda) \propto
\left \{
\begin{array}{lll}
\lambda^{1} & \mbox{if} \quad  0.001< \lambda < 0.05 & [\mu m]\\
\lambda^{-0.2} & \mbox{if} \quad  0.05 < \lambda < 0.125 & [\mu m]\\
\lambda^{-1.5} & \mbox{if} \quad  0.125< \lambda < 10.0 & [\mu m]\\ 
\lambda^{-4} & \mbox{if} \quad  \lambda>10.0  & [\mu m]
\end {array}
\right.
\end{equation}
A new model grid with a finer sampling of the parameter space and a better wavelength resolution has been created. Given that the dust density is described in polar coordinates as 
\begin{equation}
\rho\left(r,\theta \right)=\rho_0\cdot r^{-q}\cdot e^{-\gamma\times \lvert \cos(\theta)\rvert}
\label{eq:rho}
\end{equation}
the following parameters and their respective values are explored:
\begin{description}
\item[--] the torus amplitude, defined as the angular region occupied by the torus dust, complementary to the opening angle of the torus, $\Theta$: $60^{\circ}$, $100^{\circ}$ and $140^{\circ}$;
\item[--] the parameters of the dust distribution, namely $q$ : 0.00, 0.25, 0.50, 0.75 and 1.0; and $\gamma$: 0.0, 2.0, 4.0 and 6.0;
\item[--] the equatorial optical depth at 9.7 \mums, $\tau_{eq}(9.7)$: 0.1, 0.3, 0.6, 1.0, 2.0, 3.0, 6.0 and 10.0;
\item[--] the outer-to-inner radius ratio, $Y$: 10, 30, 60, 100 and 150.
\end{description}  

The model SEDs are computed at different lines of sight with respect to the torus equatorial plane in order to account for both type 1 and type 2 object emission, from 0$^{\circ}$ to 90$^{\circ}$ in steps of 10$^{\circ}$.

\subsection{Clumpy dust distribution}\label{sec:clumpy}

The N08 clumpy models, 
extensively used in the literature  \citep[see e.g.][]{mor09,nikutta09,deo11,ramos11,esquej12},
assume a grain composition of 53\% silicate and 47\% graphite, with the optical constants for the  graphite taken  from \cite{draine03} and that of the silicates from \cite{ossenkopf92}.
The primary source is described by a piecewise power-law distribution following \cite{rr95}, that when expressed in terms of  L$(\lambda)$  it takes the form:
\begin{equation}\label{eqn:pwl_nenk}
         L(\lambda) \propto
          \left \{
          \begin{array}{lll}
          \lambda^{0.2} & \mbox{if} \quad  \lambda \leq 0.01 & [\mu m] \\
          \lambda^{-1} & \mbox{if} \quad  0.01 < \lambda \leq 0.1 & [\mu m] \\
          \lambda^{-1.5}& \mbox{if} \quad  0.1 < \lambda \leq 1 & [\mu m]\\
          \lambda^{-4} & \mbox{if} \quad  \lambda >1 & [\mu m]
          \end {array}
          \right.
\end{equation}

\noindent
The angular distribution of clumps is a Gaussian of width $\sigma$, given by:
\begin{equation}
N_{T}(\beta)=N_{0}\mbox{e}^{(-\beta^{2}/\sigma^{2})},
\label{eq:N}
\end{equation}
where $\beta$ (=90-$\theta$ using the F06 notation) is the angle with respect to the torus axis, $N_{0}$ is the average number of clouds along radial equatorial rays, where the clouds follow a Poisson distribution.
Model parameters and their respective values can be summarised as follows:
\begin{description}
\item[--] the width of a Gaussian angular distribution of the clouds, $\sigma$, ranging from 15$^{\circ}$ to 70$^{\circ}$, determining the spatial distribution;
\item[--] the outer-to-inner radius ratio of the cloud distribution, $Y$= 5, 10, 20, 30, 40, 50, 60, 70, 80, 90, 100, 150, and 200;
\item[--] the average number of clouds along a given radial direction within the equatorial plane, $N_{0}$, taking values between 1 and 15;
\item[--] the power-law index determining the radial distribution of clouds, $q$ between 0.0 and 3.0;
\item[--] the optical depth of a single cloud $\tau_{V}=$5.0, 10.0, 20.0, 30.0, 40.0, 60.0, 80.0, 100.0, 150.0, calculated in the $V$ band at 0.55 \mums.
\end{description}

Again, the SEDs are created for different viewing directions with respect to the equatorial plane, from $\beta=0^{\circ}$ to $\beta=90^{\circ}$, in steps of 10$^{\circ}$.

\subsection{Intrinsic differences between the two dust distributions}\label{sec:diff}

\begin{figure}
\begin{center}
\includegraphics[angle=270,width=9cm]{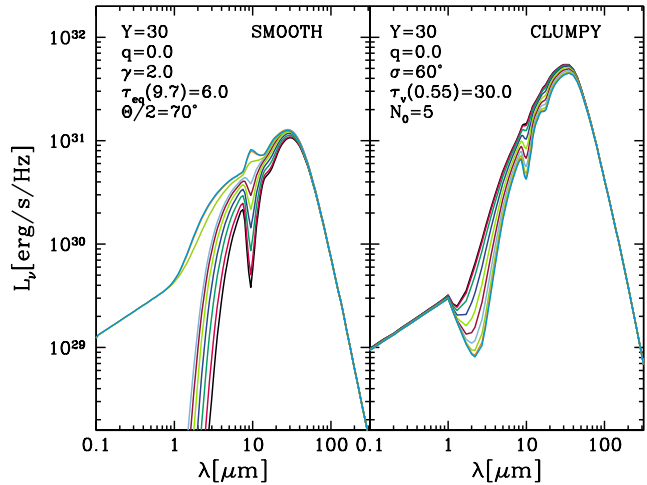}\\
\end{center}
\caption{Example of smooth (left) and clumpy (right) model SEDs with comparable model parameter values, viewed at different inclinations.}
\label{fig:modelexample}
\end{figure}

The two types of models, though both able to reproduce a variety of observations, are intrinsically quite different. Fig. \ref{fig:modelexample} shows an example of a smooth (left) and a clumpy (right) model SEDs with matched model parameters (see Sec. \ref{sec:common}), viewed at different angles ranging from 0$^{\circ}$ to 90$^{\circ}$ degrees from the equatorial plane. The sudden jump of the smooth SEDs occurs at the angle where the dust starts intercepting the line of sight (70$^{\circ}$ in this particular case). The clumpy SEDs present a smooth transition from one viewing angle to the next, due to their Gaussian angular distribution.

As already mentioned, while the F06 models use the silicates absorption and scattering coefficients given in \cite{laor93}, N08 make use of the values given in \cite{ossenkopf92}. And while the absorption coefficient for the former peaks at $\lambda \sim 9.5$ \mums, that of the latter peaks at 10  \mums. In order to keep the notation of the paper simple but also consistent with other works in the literature, we will refer to the ``9.7 \mum silicate feature'' throughout but the reader should keep this inaccuracy in mind.

The most important implications of the AGN unified scheme, is that type 1 objects are observed along a dust free line of sight, while in type 2 objects the dust intercepts the view to the nucleus. If the dust has a continuous distribution the chance of seeing the central source is a mere question of the line of sight. In a clumpy medium the chance to have a direct view of the central source is in fact the probability to encounter, on average, zero clouds, that is less than 1 even for edge-on lines of sight. To simplify the study, we only consider the two extreme inclinations i.e. edge on ($\theta=90^{\circ}$,  Eq. \ref{eq:rho}) and face-on ($\theta=0^{\circ}$) for smooth models, while we consider as type 1 and 2 clumpy models those with a probability greater and lower than 0.5, respectively, to see directly the AGN, for each model parameters combination.

\subsection{The matched model parameter spaces}\label{sec:common}

The two original model grids cover a large parameter space and produce SEDs with a considerable overlap when their parameters are ``matched''. The selection of the matched parameter space is of interest here, as is not always straight forward given the different nature of the two models. However good analogies can be found, under a few reasonable assumptions. The inner-to-outer radius ratio, $Y$, is a parameter in common between the two models. The radial variation of the dust density or that of the clouds, $q$, can also be considered as equivalent between the two models. The relation between the optical depth at 9.7 \mum and that in the V band is given by N08 as $\tau_{9.7}=0.042 \times \tau_{V}$. The equatorial optical depth $\tau_{9.7}$ for clumpy models is then equal to $0.042 \times \tau_{V} \times N_{0}$. The resulting $\tau_{9.7}$ are not identical to those used in the smooth models, the values, however, are very close and the differences they introduce in the resulting SEDs negligible. 
Finally, considering only the two extreme lines of sights, the torus opening angle is of no relevance and $\gamma$ (Eq. \ref{eq:rho}) and $\sigma$ (Eq. \ref{eq:N}) can be taken such that the distribution of dust and clumps match each other (see Table 1 for the exact values).

\begin{figure}
\centerline{
\includegraphics[angle=270,width=9cm]{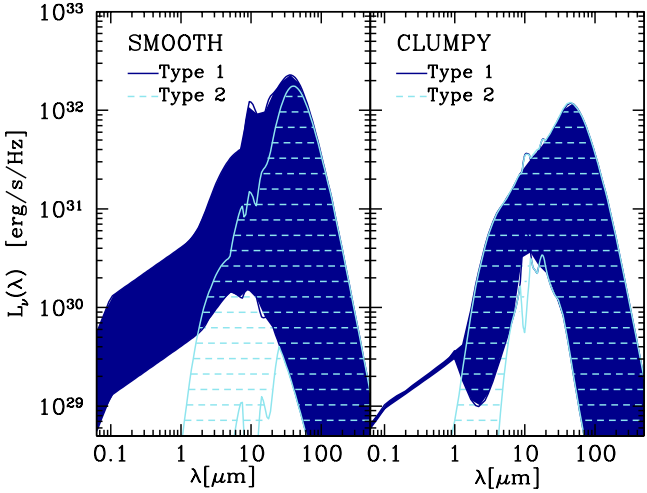}}
\caption{The range of SEDs covered by the smooth (left) and clumpy 
(right) dust configurations in the restricted parameter grids. 
The coverage is shown for both type 1 (filled regions) and type 2 (dashed) inclinations.}
\label{fig:modelrange}
\end{figure}

Table \ref{tab:parameters} summarizes the values of the models parameters that will be considered  henceforth. Fig. \ref{fig:modelrange} shows, for illustration purposes, the shapes of the model SEDs characterised by the above parameters, for type 1 and type 2 views. 

\begin{table}
\begin{tabular}{|c|c|c|}
\hline
 & SMOOTH & CLUMPY \\
\hline
$Y$ &  10, 30, 60, 100, 150 & 10, 30, 60, 100, 150 \\
\hline
$q$ & 0, 1 & 0, 1\\
\hline
$\gamma$  & 2, 4, 6 & \\
$\sigma$   & & 60$^{\circ}$, 45$^{\circ}$, 35$^{\circ}$ \\
\hline
$\tau_{9.7}$ & 0.3, 0.6, 1, 2, 3, 6, 10 & \\
$N_{0}$  & & 1 - 15\\
$\tau_{V}$  & & 5, 10, 20, 30, 40,\\
 & & 60, 80\\
\hline
\end{tabular}
\caption{Matched model parameter considered in the comparative study.}
\label{tab:parameters}
\end{table}

In matching the models parameters as described above we significantly restrict the model grids and end up comparing a total of 614 smooth and 480 clumpy models. The difference in the numbers is due to the fact that often more than one combinations of $N_{0}$ and $\tau_{V}$ correspond to the same $\tau_{9.7}$. Furthermore, since we only consider the two extreme inclinations (smooth) and probabilities (clumpy), as already explained before $\Theta$ becomes of no relevance and hence three values shown in Table \ref{tab:parameters} are consider for each given $\gamma$ and $\sigma$. 

Note that in matching the models parameters, some parameter values were left out from both model grids and the comparison presented here only applies to the restricted grids. In effect what is left out from these grids are clumpy models with very compact configuration ($q>1$) since there are no available equivalent smooth models, 
smooth models with very low optical depth ($\tau_{9.7} < 0.3$), as well as clumpy models with various combinations of $N_{0}$ and $\tau_{V}$ that do not correspond to any $\tau_{9.7}$ from the smooth grid.

\section{MODEL-TO-MODEL COMPARISON}\label{sec:comparison}

To compare the characteristics of the SEDs obtained from the two different models of dust distribution we restrict ourselves to the common model grid discussed in \S \ref{sec:common} and measure characteristic quantities, namely the prominence or {\it strength} of the silicate feature, $S$, the width of the IR bump, $W_{IR}$, the near-IR slope, $\alpha_{IR}$, the peak of the emission, $\lambda_{peak}$, and the monochromatic luminosity at 12.3 \mums.
We normalise all models to a common accretion luminosity of 10$^{45}$ erg/sec.

\subsection{The silicate features}\label{sec:sifeature}

The silicate emission feature at $\sim$9.7 \mum has been under the spotlights because its prominent appearance, mostly in smooth models, did not match any observations for many years, questioning the reliability of the Unified Scheme and the accuracy of the models themselves. In clumpy models, on the other hand, this feature is often smeared out in axial viewing in spite of its prominence in emission from directly individual clouds. The controversy on the prominence of the feature was partly solved by observations carried out with \textit{Spitzer}/IRS \citep[e.g.][]{siebenmorgen05,hao05,shi06}.

\begin{figure}
\centerline{
\includegraphics[width=9.5cm]{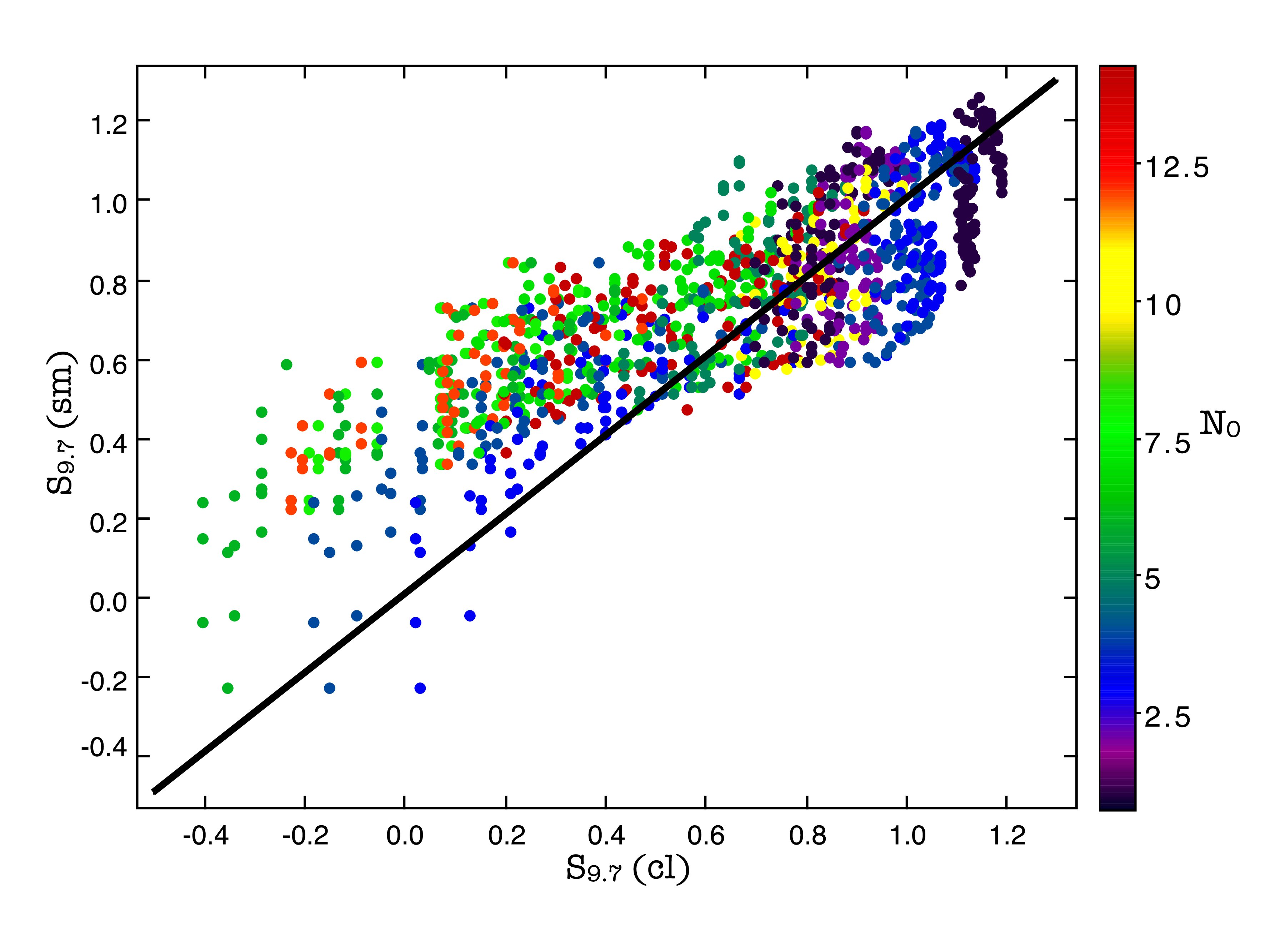}}
\caption{The values of $S_{9.7}$ for matched smooth and clumpy models. The points are colour-coded based on the value of $N_{0}.$}
\label{fig:S97type1}
\end{figure}

The strength of the silicate feature is defined as the logarithm of the ratio between the flux $F$ measured within the line profile over the continuum flux $F_{c}$ at such wavelength, i.e.
\begin{equation}
S=\mbox{ln}\left(F(\lambda_m)/F_c(\lambda_m)\right)
\label{eq:S}
\end{equation}
where $\lambda_m$ is the wavelength at which the feature's strength is an extremum with a value in the interval between 8.5 and 11.5 \mum \citep[for the computation of $F_{c}$ see][]{sirocky08}. As already mentioned in Sec. \ref{sec:diff}, F06 and N08 models consider different absorption coefficients that peak at different wavelengths, but for simplicity we call the strength of the silicate feature around 9.7 \mum $S_{9.7}$, irrespective of $\lambda_m$.

\begin{figure*}
\centerline{
\includegraphics[angle=270,width=9cm]{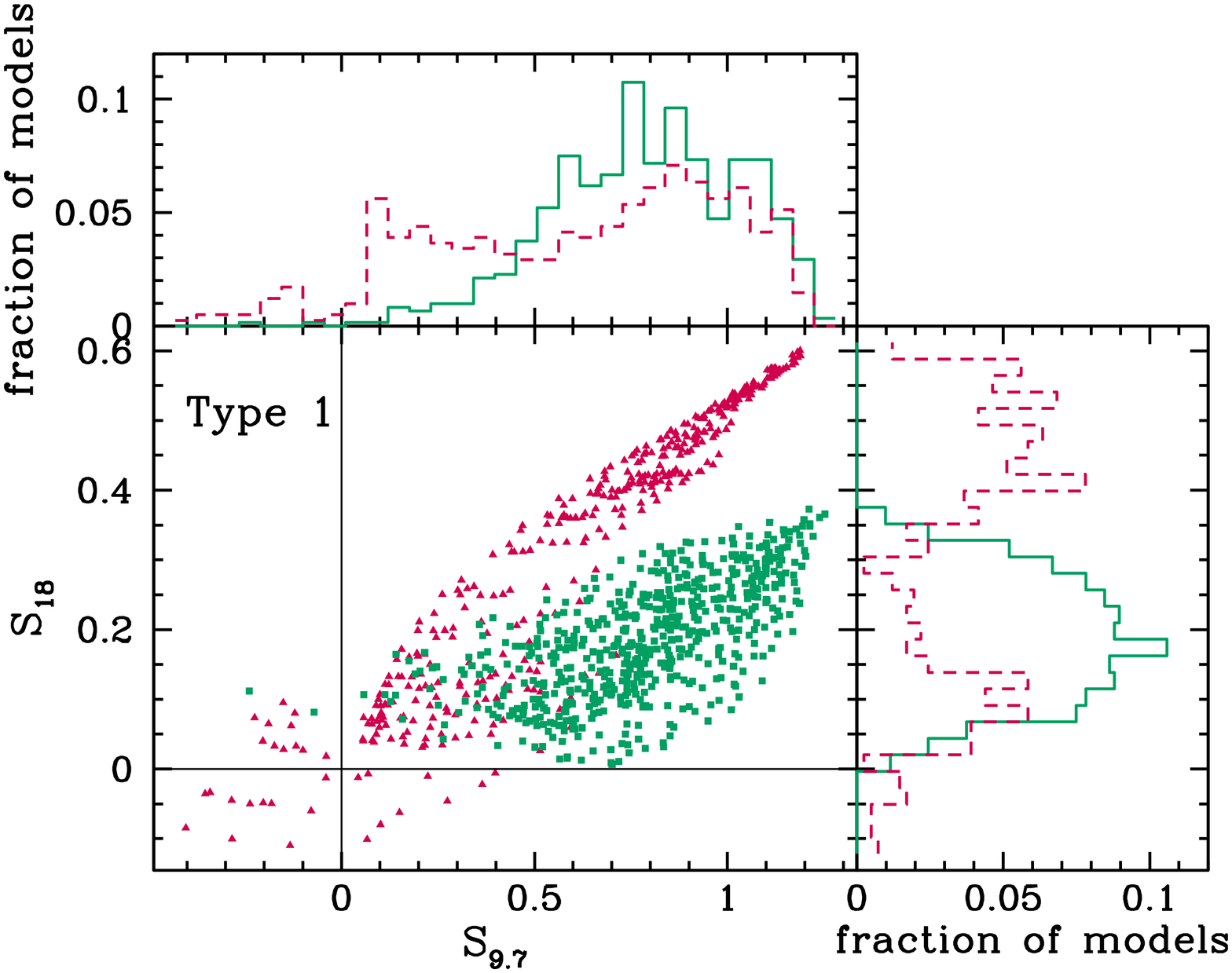}
\includegraphics[angle=270,width=9cm]{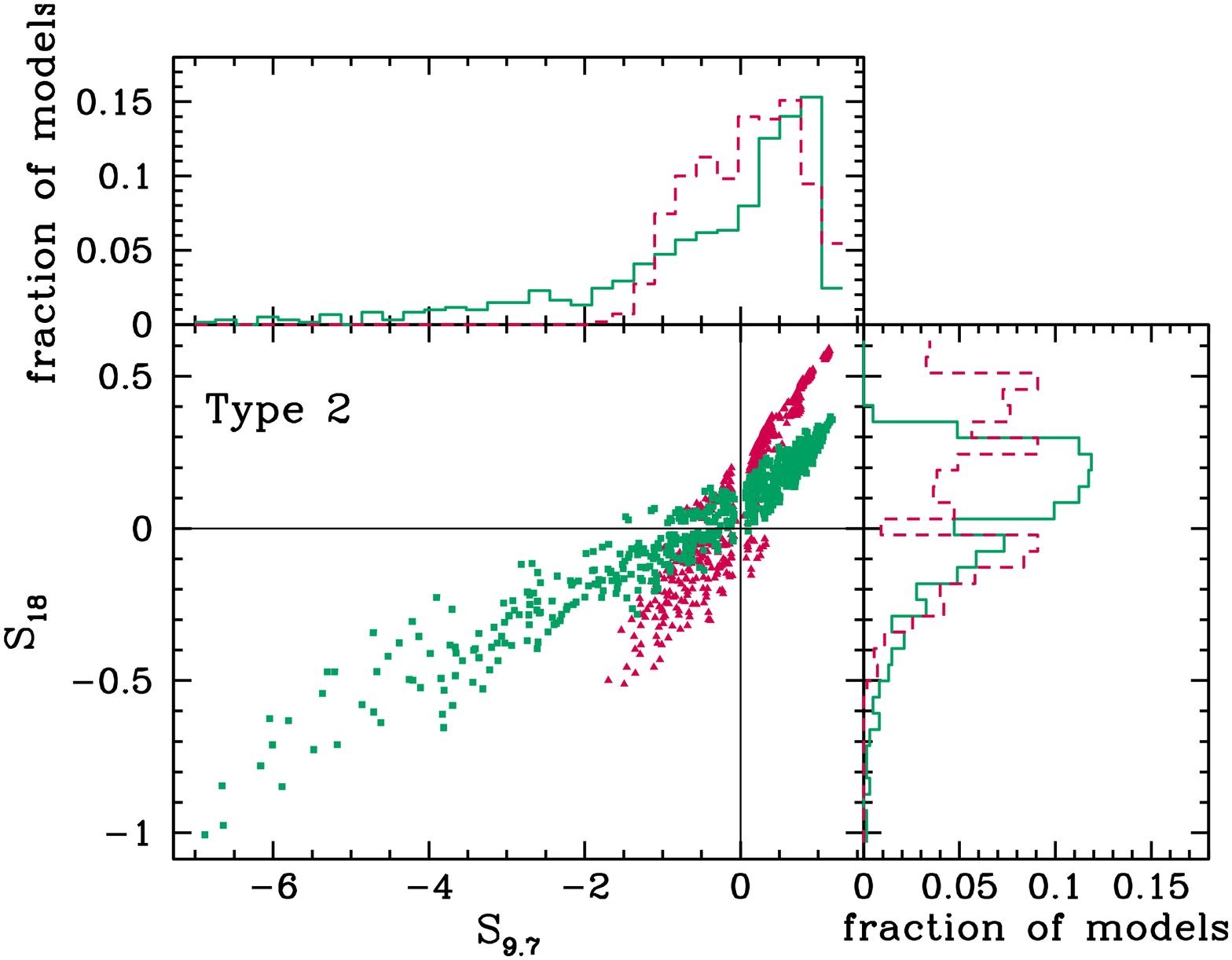}}
\caption{$S_{18}$ versus $S_{9.7}$ for smooth (green squares) and clumpy (red triangles) models,  and $S_{9.7}$ and $S_{18}$ distributions for type 1 (left) and type 2 (right) views.}
\label{fig:S}
\end{figure*}

Fig. \ref{fig:S97type1} compares the values of $S_{9.7}$ for matched smooth and clumpy models in a face-on inclination. The symbols are colour-coded according to the value of $N_{0}$. For small values of $N_{0}$ the values of the feature for the two models lie close to the 1:1 line (shown in black), however as  $N_{0}$ takes larger values the points deviate from the line, with the clumpy models showing a weaker feature. This is due to increasing attenuation with increasing number of clouds, as dust is optically thick to itself.
The distribution of $S_{9.7}$ for the two dust configurations can be seen in the two top histograms  in Fig. \ref{fig:S}, with the green (red) lines corresponding to smooth (clumpy) configurations. 

Both Fig. \ref{fig:S97type1} and the histograms at the top of Fig. \ref{fig:S} show that a large interval of $S_{9.7}$ values is covered by both dust configurations. The clumpy models with type 1 views can produce equally strong silicate features in emission (Fig. \ref{fig:S}, top left histogram), despite repeated claims in the literature to the contrary. A large fraction of them, however, does extend to weaker $S_{9.7}$ values. The striking differences in the behaviour of $S_{9.7}$ (with means of 0.79 and 0.62 for type 1 smooth and clumpy, respectively, and -0.48 and 0.06 for type 2 views) have their origin in both the different chemical compositions used by the two models (see Sec. \ref{sec:diff}) and the fact that only the restricted model grids are being compared, leaving out a number of models (both smooth and clumpy) with parameters that can not be matched by the other dust distribution.
On the other hand, the long standing issue of very deep silicate absorption produced by smooth models is confirmed, with the tail of the $S_{9.7}$ distribution for edge-on inclinations (Fig. \ref{fig:S}, top right histogram) extending to large negative values.
Both models can produce silicate emission in type 2 views. Since objects with such characteristics are rather uncommon \citep[but not unheard of, see e.g.][]{sturm06,teplitz06,mason09,nikutta09}, the models that produce such features should also be seen as non-representative, yet realistic.  Furthermore, silicate in absorption in type 1 views is also produced by both models, although only marginally.

Recently the attention has turned towards the 18 \mum silicate feature which, being broader and fainter, had almost escaped detection. 
Its strength, $S_{18}$, is also computed from Eq. \ref{eq:S} with its extremum taking values in the interval between 17.0 and 19.5 \mums. 
$S_{9.7}$ and $S_{18}$ are shown in Fig.  \ref{fig:S} for type 1 (left) and type 2 (right) views.

Clumpy models show stronger $S_{18}$ emission in type 1 views and span a larger range compared to the smooth models. For type 2 views, the values for $S_{18}$ emerging from the two dust configurations are somewhat more similar, but again clumpy models extend to stronger features in emission while there are a few smooth models that show very deep absorption.

\cite{thompson09} proposed that the comparison of the strengths of the two silicate features could be used as a diagnostic of dust composition and, also, to discriminate between smooth and clumpy distributions. Fig. \ref{fig:S} shows indeed that the relative strength of the two features is very distinct between the two models, however as pointed out by \cite{sirocky08} this difference reflects the effect of the different chemical compositions used in the two models (discussed in Sec. \ref{sec:diff}; see also  Figs. 7 and 9 in \citealt{sirocky08} showing the changing slope in the $S_{18}$ vs $S_{9.7}$ distribution as a function of the dust chemistry).

\subsection{Infrared SED}\label{sec:IRSED}

\begin{figure}
\centering{
\includegraphics[angle=270,width=8.5cm]{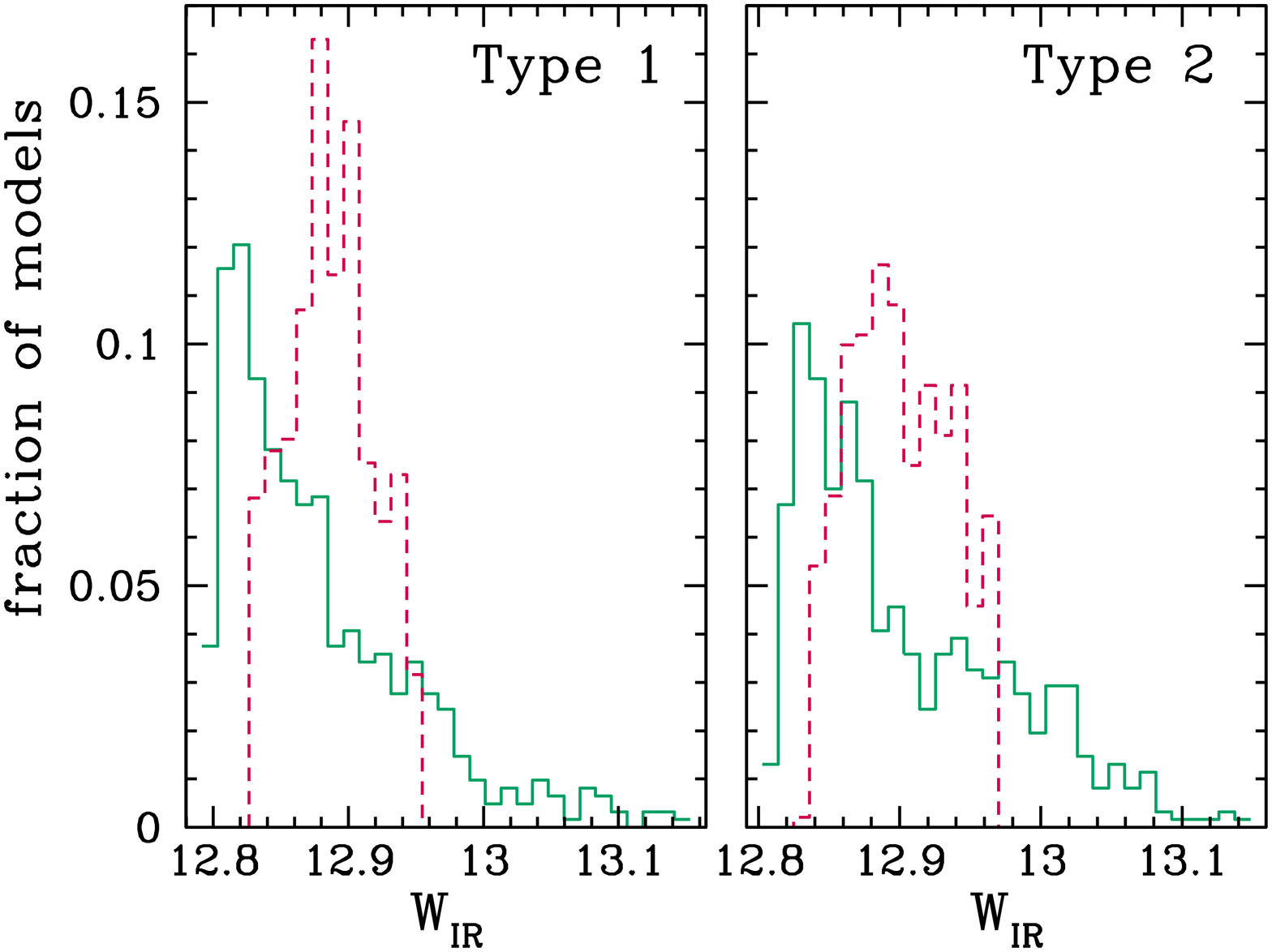}}
\centering{
\includegraphics[angle=270,width=8.5cm]{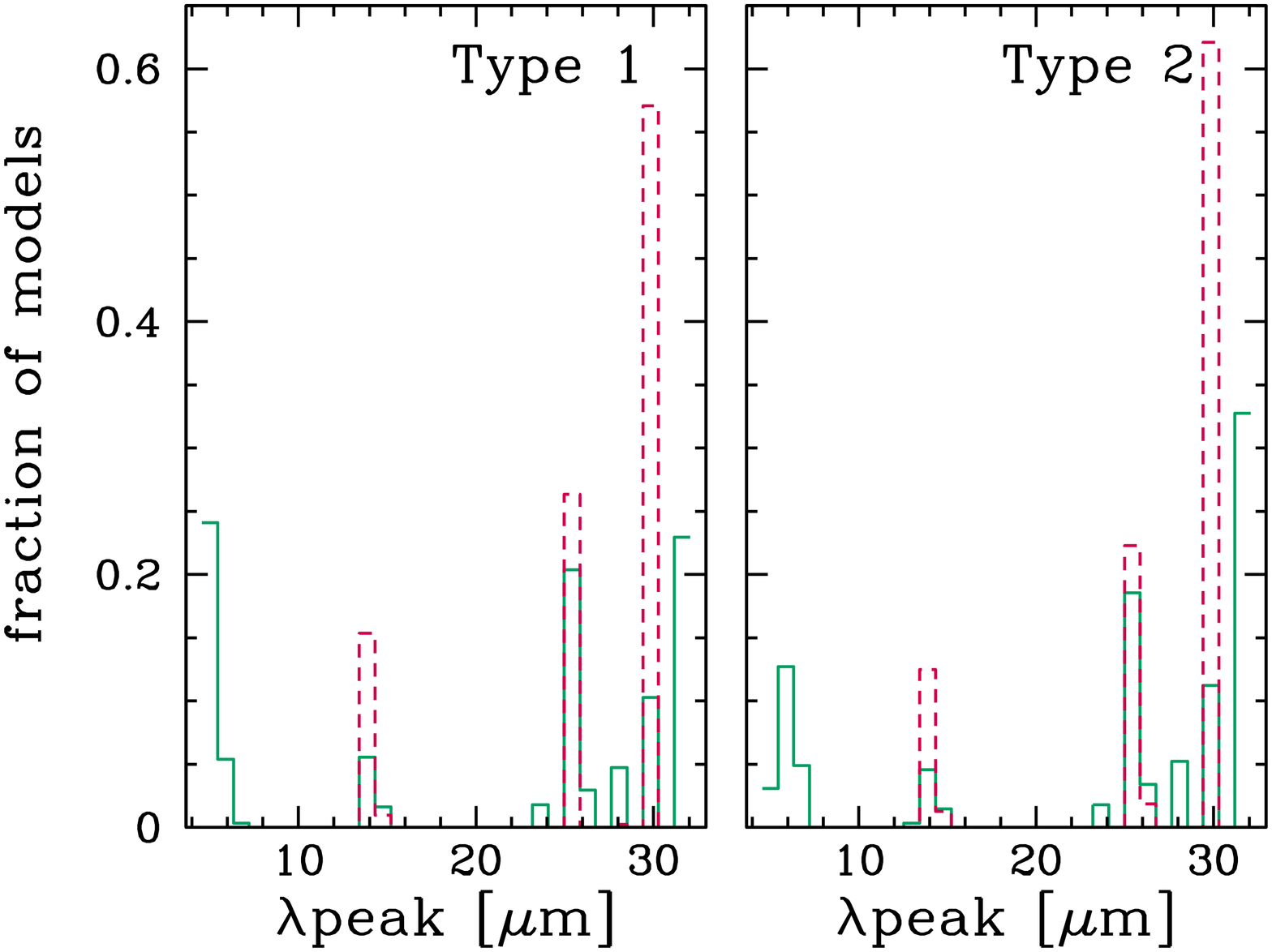}}
\centering{
\includegraphics[angle=270,width=8.5cm]{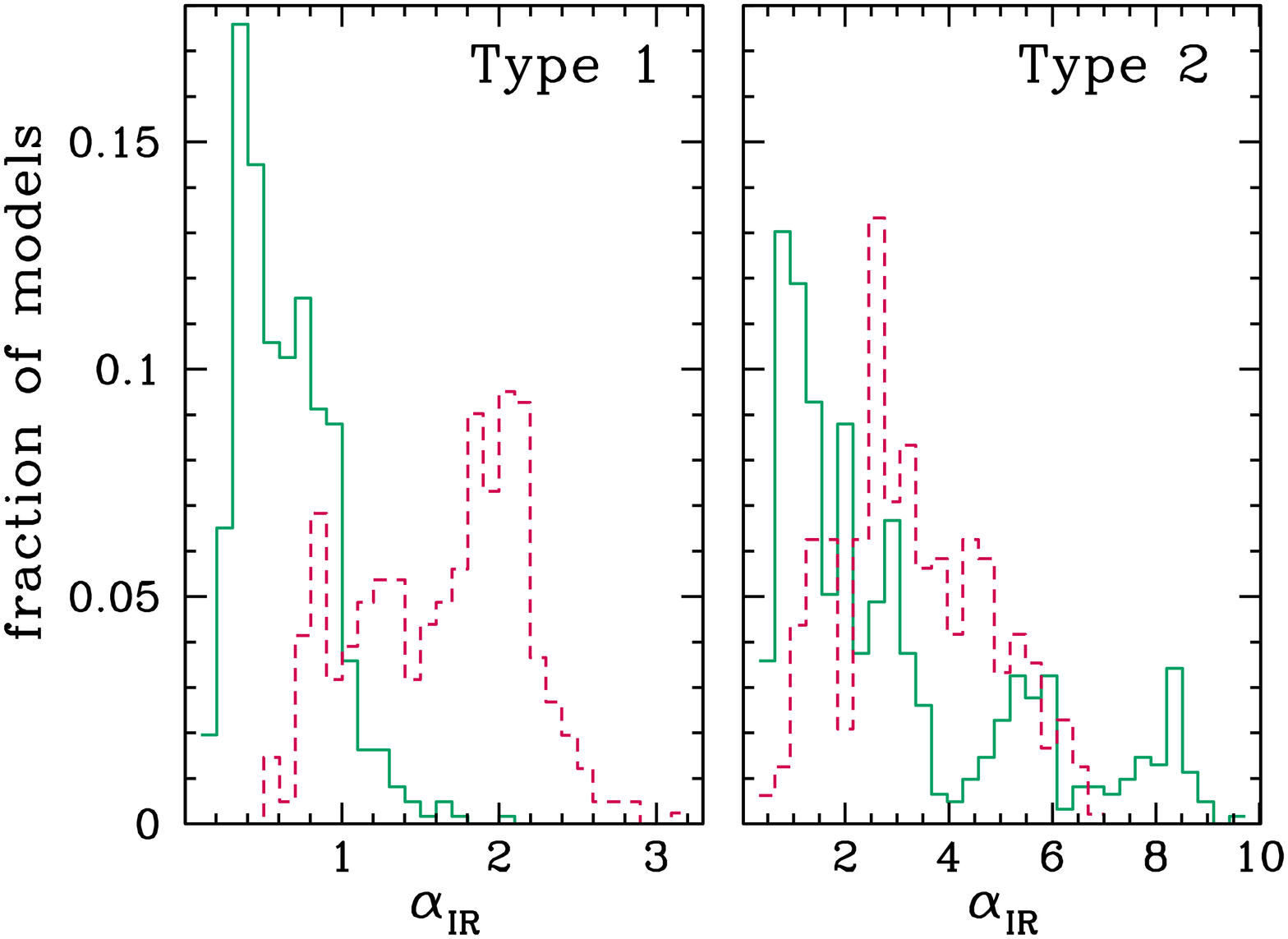}}
\caption{From top to bottom: W$_{IR}$, $\lambda_{peak}$ and  $\alpha_{IR}$, for type 1 (left column) and type 2 (right column) views, for smooth (continuous lines) and clumpy (dashed lines) models.}
\label{fig:SEDcomp}
\end{figure}

We now examine the properties of the IR model SED resulting from two different dust distributions, namely:

\begin{description}
\item[--] the width of the IR bump, W$_{IR}$, defined as the log$_{10}$ of the frequency range in which the spectrum is more than 1/3 of its peak value (expressed in $F_{\nu}$) as in e.g. 
\cite{granato94};
\item[--] the wavelength where the flux value (in $F_{\nu}$) of the SED peaks, $\lambda_{peak}$;
\item[--] the spectral index at near-IR wavelengths, defined as
	\begin{equation}
	 \alpha_{IR}=\frac{\log_{10}(F_{4.5})-\log_{10}(F_{3.5})}{\log_{10}(\lambda_{4.5})-\log_{10}(\lambda_{3.5})}
	\end{equation}
\item[--] the monochromatic luminosity at 12.3 \mums, $L_{12.3}$.
\end{description}
The distribution of the first three parameters for type 1 and 2 views are shown in Fig. \ref{fig:SEDcomp}. 

While the mean of the distributions of W$_{IR}$ is the same for smooth and clumpy models (12.88  for type 1 and 12.9 type 2 views), and the medians differing only slightly (12.86 versus 12.88 for type 1 views and 12.89 versus 12.9 for type 2 views), clumpy models produce, on average, wider IR bumps, as already noted by other authors (e.g.  \citealt{dullemond05,nenkova02}). In a smooth medium the dust temperature is a continuous and monotonic function of the distance from the central source. In a clumpy medium, instead, the non-illuminated side of the clouds, particularly those with high optical depth, will be in general much colder with respect to their illuminated side. The net result is that a whole range of temperatures can coexist within the torus at a given distance from the primary source \citep[see][]{schartmann08}. The temperature of the dust within each individual cloud spans a wide range and therefore each region of the torus emits strongly at all wavelengths, from mid-IR to submm. The rms of the W$_{IR}$ distribution of smooth models is twice as large as that of clumpy models and smooth models can also produce large $W_{IR}$. In fact, the largest ones in this study are indeed produced by the smooth models in the matched grids. 

The peak of the IR SEDs, $\lambda_{peak}$, measured on the continuum alone, excluding the silicate feature is shown in the middle panel of Fig. \ref{fig:SEDcomp}. Both sets of models peak at wavelengths typically between 10 and 30 \mums, a fraction of smooth models, however, have their peak at very short wavelengths. For both views, the majority of clumpy models culminate at $\sim$30 \mums, a behaviour related but not confined to the wider, on average, clumpy SEDs.

The distribution of the near-IR spectral index, $\alpha_{IR}$, varies a lot depending on the dust morphology (Fig. \ref{fig:SEDcomp}, lower panel). For type 1 views in particular, the  range of overlap is very small, with clumpy models producing, on average, steeper spectra. The mean (median) of the distributions for the smooth and clumpy models are 0.63 (0.58) and 1.63 (1.74), respectively. The same occurs for type 2 views but to a lesser extent, with a mean (median) of 2.98 (2.1) and 3.32 (3.18) for smooth and clumpy dust, respectively. Three components contribute to these differences: i) the lack of a very hot component in the clumpy models (see e.g. \citealt{deo11}), ii) the primary 
source (different power laws and different wavelength coverage for the two dust distributions), and, to a much lesser extent, iii) the scattering in the clumpy medium, which strongly depends on the distribution of the clouds. We calculated the average fractional contribution of the primary source to the flux at 3.6 and 4.5 \mum to be of 1.5\% and 0.7\%, respectively, for clumpy models, while it is constant and $\sim$44\% in smooth models in both bands. This would account for about 30\% of the difference in the values of $\alpha_{IR}$ between the two dust configurations and, therefore, the remaining difference must be due to point i.

Some recent works indicate that the mid-IR emission, in particular the monochromatic luminosity at 12.3 \mums, $L_{12.3}$, can be used as a diagnostic to distinguish between a smooth and a clumpy configuration. \cite{horst06} found that in a sample of eight Seyfert galaxies (a Seyfert 1, an 1.2 and six 1.5 or later), $L_{12.3}$ is tightly correlated with their X-ray $L_{2-10} keV$ luminosity, regardless of their type. This was interpreted as an evidence for the dust being optically thin at 12.3 \mums, a characteristic which was reported to be typical of clumpy models, but incompatible with the smooth model of \cite{pier92}.

\begin{figure}
\centerline{
\includegraphics[angle=270,width=9cm]{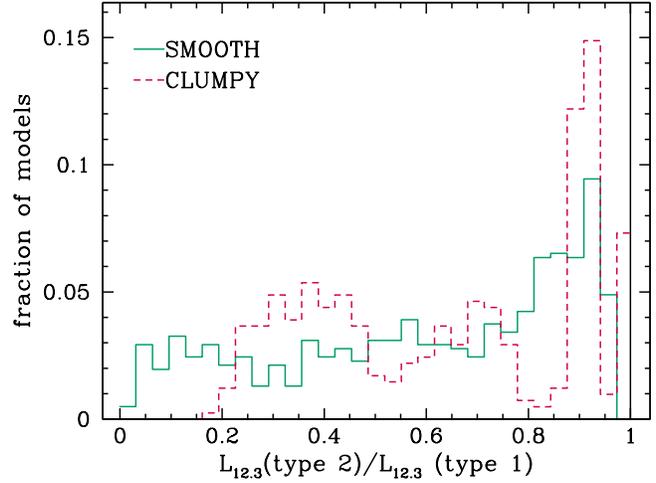}}
\caption{The distribution of the ratio of $L_{12.3}$ in type 2 over type 1 views, for smooth and clumpy models.}
\label{fig:L12}
\end{figure}

We have checked the sets of clumpy and smooth models against this prediction, by computing the   ratio of $L_{12.3}$ in type 2 over type 1 views, shown in Fig.  \ref{fig:L12}. With the exception of 7\% of clumpy models that are very close to be optically thin, i.e. with $L_{12.3}^{type_2}/L_{12.3}^{type_1} \sim 1$, both sets of models always lie well below this value. Overall, although the shape of the two distributions is different, we find no real evidence to the above claim.

We have shown that even models with matched parameters can produce very different features (see Fig. \ref{fig:S97type1}). But could random parameter combinations result in very similar model SEDs? In order to answer this question, we introduce a measure of the dissimilarity of any two SEDs based on the properties discussed in the previous sections as follows:

\begin{align}\label{eq:delta}
\bar\Delta&=\frac{1}{6} \left( \left|\frac{S_{9.7}^{sm}-S_{9.7}^{cl}}{S_{9.7}^{sm}}\right|+\left|\frac{S_{18}^{sm}-S_{18}^{cl}}{S_{18}^{sm}}\right|+\left|\frac{W_{IR}^{sm}-W_{IR}^{cl}}{W_{IR}^{sm}}\right| \right.
\nonumber\\
              & \left. + \left|\frac{\lambda_{peak}^{sm}-\lambda_{peak}^{cl}}{\lambda_{peak}^{sm}}\right| + \left|\frac{\alpha_{IR}^{sm}-\alpha_{IR}^{cl}}{\alpha_{IR}^{sm}}\right| + \left|\frac{C^{sm}-C^{cl}}{C^{sm}} \right| \right)
\end{align}

\noindent 
where $sm$ and $cl$ denote  smooth and clumpy models, respectively, $C=F_{24}/F_{4.6}$ and $F_{4.6}$ and $F_{24}$ are the fluxes at 4.6 and 24 $\mu$m respectively. $C$ has been introduced to take into account the relative fluxes of the (normalised) models in addition to the IR spectral index, $\alpha_{IR}$.
Comparing the value of $\bar\Delta$ for each pair of matched models, we find that none of the pairs have a value lower than 0.1. Furthermore, we only find 6 combinations for type 1 views and 21 combinations for type 2 views with $\bar\Delta \le 0.1$, with very different model parameters. We therefore conclude that there are virtually no random pairs of smooth and clumpy models in the matched grids that can produce very similar SEDs.

\section{DISCUSSION}\label{sec:discussion}

Within the paradigm of the AGN unified scheme, the different properties of the various types of AGN are  attributed to the interception of the line of sight by an axisymmetric molecular dust distribution encircling the nucleus, likely in the form of a torus. Direct and indirect observational evidence, such as the dichotomy of the AGN population in obscured and unobscured objects, the ionization cones for the NLR, the MIR bump present in the SEDs of all known AGN attributed to emission by dust heated by the primary source, but also IR interferometry that directly resolved the inner parsec of the nucleus of the prototype Seyfert 2 galaxy NGC 1068 \citep{wittkowski04, jaffe04}, all support the existence of a toroidal absorbing structure. As there is evidence favouring both a clumpy distribution \citep[e.g.][]{risaliti02, wittkowski04,tristram07} and a smooth distribution \citep[e.g.][]{ibar07}, both smooth and clumpy models are still equally widely used to explain the observed SEDs of AGN, counting both successes and shortcomings.

In this paper we compared two sets of models widely used in the literature, each representative of one of the two dust distributions, namely an updated grid of the smooth models by \cite{fritz06} and the clumpy models by \cite{nenkova08a}. In order to compare as similar models as possible and despite the intrinsic differences between the two configurations, we matched the two sets of parameters building thus restricted grids, and compared only the models within these grids. This approach has the limitation of not exploring the full parameter space for either of the two models. Furthermore, and to avoid complications arising from the probabilistic nature of viewing an AGN as a type 2 object through a clumpy medium, we consider two extreme inclinations i.e. edge on and face-on, for smooth models, and take as equivalent type 1 and 2 clumpy models those with a probability to see directly the central engine greater and lower than 0.5, respectively. Our findings can be summarised as follows: 

\begin{itemize}
\item[--] Even after matching the model parameters there is not a one-to-one correspondence. For each smooth model of the restricted grid there can be several clumpy models with matched $Y$, $q$, and $\sigma$, but different combinations of $N_{0}$ and $\tau_V$ that finally produce a given value of $\tau_{9.7}$. Additionally, and since we only consider the two extreme inclinations (face-on and edge-on), the torus opening angle can not be matched and for each clumpy model we consider all smooth counterparts, irrespective of the value of $\Theta$.

\item[--] The distribution of the various features of the IR model SEDs differs when smooth and clumpy dust configurations are considered. The very different behaviour of the silicate features is due more to the different chemical compositions assumed by F06 and N08 and less to the actual dust morphology. We confirm the occurrence of broader, on average, IR SEDs in clumpy configurations, with a larger fraction of clumpy models peaking at long wavelengths ($\sim$30 \mums). The infrared spectral index, especially for type 1 views, is the quantity that changes the most between the two dust configurations, owing to the differences in the primary source assumed in the two models as well as the lack of the hotter dust component in the treatment of the clumpy medium.

\item[--] Our study showed no large differences in the behaviour of $L_{12.3}^{type_2}/L_{12.3}^{type_1}$, as an indicator of how optically thin a medium is, between the two model configurations, with 5\% of clumpy models showing a value very close to 1 (i.e. optically thin configuration), a similar amount of clumpy and smooth models ($\sim$ 20\%) having $L_{12.3}^{type_2}/L_{12.3}^{type_1} \ge 0.9$ and the rest of both dust configurations lying below this value. 

\item[--] Models with matched parameters within the restricted model girds do not produce similar SEDs (similar either by eye or based on the value of $\bar\Delta$ introduced in \S \ref{sec:IRSED}). Additionally, only a very limited number of random parameter combinations can result in seemingly identical SEDs, though the dust configuration differs. 
\end{itemize}

From the above we conclude that, even though the two dust models produce distinct SEDs, most of the differences arise from the model assumptions (e.g. primary source, dust chemical composition) and not from the dust morphology (smooth or clumpy).
To summarise, the properties of dust in AGN as measured by matching observations (be it broad band IR photometry or IR spectra) with models will strongly depend on the choice of the dust distribution. The possibility to discriminate between a smooth and a clumpy medium based on the various SED features may exist, but ambiguities are more common than not.
Independent estimates of physical parameters, such as the optical depth, the size of the torus or the mass of the gas are needed in order to further constrain the models. X-ray observations could, for instance, provide an {\it upper limit} of the optical depth, integrated along the line of sight, high resolution HI maps of known nearby AGN could put constraints on the gas content within the circumnuclear region.

Eventually, ALMA will permit to indirectly determine the morphology of the obscuring material by allowing the comparison between the mid-IR and sub-mm emission of the structure (see e.g. \citealt{maiolino08}) or even to directly resolve the obscuring torus, making use of its very high angular resolution (sub-pc scale at the distances of nearby AGN at high frequencies).

\section*{ACKNOWLEDGMENTS}
 
This work makes use of the \cite{nenkova08a,nenkova08b} models: http://www.pa.uky.edu/clumpy/. A.F. would like to thank Robert Nikutta and Moshe Elitzur for providing detailed explanation on their models and on the calculation of the output parameters, as well as Hagai Netzer for very useful comments. We also thank R. Nikutta for his details and insightful referee report, that improved a lot the content and quality of our work. This work makes use of TOPCAT, developed by M. Taylor.

\label{lastpage}

\end{document}